\documentclass[reprint, prx, aps, notitlepage, nofootinbib,longbibliography,floatfix, superscriptaddress]{revtex4-1}
\usepackage{amsmath,amstext,amssymb,graphicx,bm,dcolumn,color,times, braket,caption}
\usepackage[colorlinks=true,citecolor=blue,urlcolor=red]{hyperref}
\usepackage[table]{xcolor}
\usepackage{graphicx}
\usepackage{amssymb}
\usepackage{amsmath}
\usepackage{amsthm}
\usepackage{amsfonts}
\usepackage{braket}
\usepackage{lineno}
\usepackage{subfig}
\usepackage{physics}
\begin{document}

\title{Optimizing the phase sensitivity of a Michelson interferometer with a two mode squeezed coherent input}

\author{Stav Haldar}
\affiliation{Hearne Institute for Theoretical Physics, Department of Physics and Astronomy, Louisiana State University, Baton Rouge, Louisiana, 70803, USA}
\email{hstav1@lsu.edu}

\author{Pratik J. Barge}
\affiliation{Hearne Institute for Theoretical Physics, Department of Physics and Astronomy, Louisiana State University, Baton Rouge, Louisiana, 70803, USA}

\author{Xiao-Qi Xiao}
\affiliation{Department of Communication Engineering, Shanghai Dianji University, Shanghai 200240, China}

\author{Hwang Lee}
\affiliation{Hearne Institute for Theoretical Physics, Department of Physics and Astronomy, Louisiana State University, Baton Rouge, Louisiana, 70803, USA}

\begin{abstract}
A Michelson-type interferometer with two-mode squeezed coherent state input is considered. Such an interferometer has a better phase sensitivity over the shot-noise limit by a factor of $e^{2r}$, where $r$ is the squeezing parameter [Phys. Rev. A 102,022614 (2020)]. We show that when photon loss and noise in the two arms is asymmetric an optimal choice of the squeezing angle can allow improvement in phase sensitivity without any increase in input or pump power. In particular, when loss occurs only in one arm of the interferometer, we can have improvement in phase sensitivity for photon loss up to 80\%. Hence, a significant improvement can be made in several applications such as LiDAR, gyroscopes and measuring refractive indices of highly absorptive/reflective materials.
\end{abstract}

\maketitle


\section{Introduction}
Quantum enhancement of phase estimation using optical interferometers is an active area of research. The phase sensitivity of an interferometer using an ordinary coherent light source scales as $1/\sqrt{n}$, where $n$ is the mean number of input photons. This scaling limit which is due to the photon counting error is called the shot-noise limit \cite{kok2012}. Over the last four decades, a lot of efforts have been made to overcome this limit. Largely, there are three distinct approaches depending on whether it uses squeezed states, photon-number states, or some combination of both.
The first one, squeezed-state approach, is to combine the ordinary coherent light with squeezed state at the first beam splitter. It is the scheme that was proposed by Caves for gravitational wave detection in the early 1980’s \cite{caves1981}. The phase sensitivity is shown to scale as $e^{-r}/\sqrt{n}$ or higher under certain conditions \cite{pezze2008, seshadreesan2011, pooser2019} with the squeezing parameter $r$. SU(1,1) interferometers introduced by YMK, where the usual beam splitters are replaced by four-wave mixers \cite{YMK1986}, its coherently-boosted scheme \cite{plick2010,marino2012,hudelist2014}, and the two-mode squeezed-vacuum scheme \cite{anisimov2010} can also be included in this category.
The second one, number-state approach, is typically to use fixed number of photons distributed to two input ports of the interferometer. Such an approach was first proposed by Yuen \cite{Yuen1986} and YMK \cite{YMK1986} in the 1980’s. Many different correlations between the two-mode number states were proposed to go beyond the shot-noise limit \cite{holland1993, kim1998, kuzmich1998, berry2000, campos2003, pezze2006, lee2009}. The phase sensitivity in this case typically scales as $1/n$, dubbed as the Heisenberg limit \cite{higgins2007, nagata2007}. Dual- Fock, or twin-Fock, states \cite{holland1993, kim1998, kuzmich1998, campos2003}, intelligent states \cite{hillery1993, brif1996}, and N00N states \cite{lee2002, resch2007, afek2010} are among the named correlated Fock states.
The third category is the approach that combines the squeezed states and the number states. This approach was first proposed by Gerry et al.
\cite{birrittella2012, carranza2012}. More often than not the input state to the interferometer is prepared by an operation---such as photon addition, subtraction, or catalysis---made onto squeezed states to achieve quantum enhancement in phase sensitivity \cite{zhang2021, kumar2022}. These approaches might as well be differentiated as gaussian states, non-gaussian states, and non-gaussian operation on gaussian states in more general terms.
It has been shown in a recent paper that a coherently boosted two-mode squeezed state, or two-mode squeezed coherent state (TMSCS) can be used as the interferometer input to achieve sub-shot-noise phase sensitivity \cite{xiao2020}. The interferometer is a generic, ordinary SU(2) type with two beam splitters and intensity difference measurement at the output---as opposed to the seeded SU(1,1) type. The phase sensitivity in this case is shown to be $e^{-2r}/\sqrt{n}$. We have a doubly enhanced phase sensitivity; one with squeezing, the other with amplification. Thus, the value of $n$ is $e^{2r}$ times larger than the number of photons in the initial coherent states. 

In the present work we consider a Michelson-type interferometer with TMSCS as the input. First in Section \ref{sec:phase_sensitivity}, we investigate the maximum amount of loss and noise tolerable to maintain sensitivity below the shot-noise limit. In particular, we report that by an optimal choice of the squeezing angle or other input phases, the phase sensitivity can be enhanced and noise beyond the 3 dB limit tolerated. Next, in Section \ref{sec:radiation_pressure}, we analyze the complementary radiation-pressure error. We find that the radiation pressure noise increases with decreasing photon counting fluctuations, so that the standard quantum limit remains intact. In Section \ref{sec:conclusions} we discuss our conclusions.
\section{Phase Sensitivity of a Michelson Interferometer}
\label{sec:phase_sensitivity}
\begin{figure*}[ht]
\includegraphics[width=16cm]{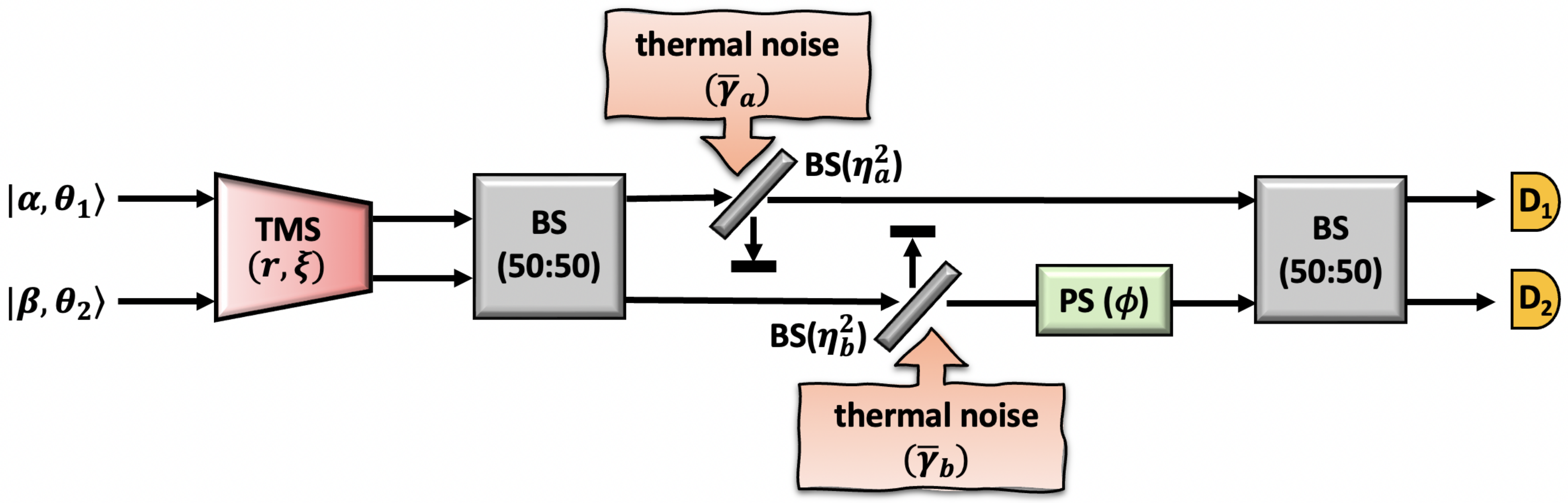}
\caption{\label{fig:michelson} 
TMS: two mode squeezer, BS: beam splitter, PS: phase shifter, D$_{1,2}$: detectors. We consider that the two arms have transmissivity $\eta_a^2$ and $\eta_b^2$, respectively and thermal noise with mean photon numbers $\bar{\gamma}_a$ and $\bar{\gamma}_b$, respectively. A two mode squeezed coherent source is used as the input to the interferometer, created by passing two coherent sources with average photon numbers $\alpha, \beta$ and phases $\theta_1, \theta_2$ through an optical parametric amplifier with squeezing factor $r$ and squeezing angle $\xi$. The two 50:50 BS operations shown in this schematic both occur at the sole beam splitter of the MI. The loss parameters $\eta_{a,b}^2$ instead of $\eta_{a,b}$ have been used because for the MI photons in each mode traverse their respective arms twice before reaching the detectors D1 and D2.}
\end{figure*}

Here we describe the phase sensitivity of a Michelson interferometer (MI) with loss and noise in both arms (See Fig.~\ref{fig:michelson}). The relative phase $\phi$, acquired by the photons in one arm (variable arm) compared to the other arm (reference arm) can be measured by its modulation of the photon number difference at the two outputs of the interferometer (D1 and D2 in Fig.~\ref{fig:michelson}). This can in turn be used to estimate the path difference between the two arms. The higher the sensitivity of the photon number difference to the changes in relative phase the more precise the interferometer. Here, in order to model a realistic setting we assume loss and noise in both the reference arm and variable arm of the interferometer.

Let $\ket{\Psi_{in}}_{ab}$ be the two-mode ($a$ and $b$) input state for the interferometer. The output state going into the two detectors D1 and D2 is then be given by:
\begin{eqnarray}
    \ket{\psi_{out}}_{ab} &=&\\ \nonumber
    \hat{BS}_{ab}&\hat{BS}_b&(\eta_b^2)\hat{BS}_a(\eta_a^2)\hat{PS}_{ab}\hat{BS}_{ab} \ket{\Psi_{in}}_{ab} ,
\end{eqnarray}
that is, the evolution of the state $\ket{\Psi_{in}}_{ab}$ through the interferometer is given by the the operations of a 50:50 beam splitter -- $\hat{BS}_{ab}$, a relative phase-shift between the two arms -- $\hat{PS}_{ab}$, beam splitter operators modelling the loss and noise on both arms -- $\hat{BS}_a(\eta_a^2)$ and $\hat{BS}_b(\eta_b^2)$. Finally, another operation of the 50:50 beam splitter recombines two beams.

The operators mentioned above are defined in the following way, by their actions on the two modes $\hat{a}$ and $\hat{b}$ of the interferometer.

50:50 beam-splitter transformation:
\begin{eqnarray}
\hat{BS}_{ab}^\dagger \begin{bmatrix}
\hat{a}' \\
\hat{b}'
\end{bmatrix}\hat{BS}_{ab} = \frac{1}{\sqrt{2}}\begin{bmatrix} 1 & i \\
i & 1\end{bmatrix}\begin{bmatrix}
\hat{a} \\
\hat{b}
\end{bmatrix} = \begin{bmatrix}
\hat{a} + i\hat{b} \\
i\hat{a} + \hat{b}
\end{bmatrix}. 
\label{eqn:bs}
\end{eqnarray}
Beam-splitter transformations for loss and noise modelling act on one mode $a$ or $b$ of the interferometer and one thermal mode $c$ or $d$. The loss is characterised by the transmissivity of the beam-splitter $\eta_a^2$ or $\eta_b^2$ (higher transmissivity means lower loss). $\eta_{a,b}^2$ instead of $\eta_{a,b}$ have been used because for the MI photons in each mode traverse their respective arms twice before reaching the detectors D1 and D2. The noise is characterized by the mean photon numbers $\bar{\gamma}_{a,b}$ of the external modes $c,d$. These external modes will eventually be traced out (not measured). These operations are given by:
\begin{eqnarray}
\hat{BS}_a^\dagger \begin{bmatrix}
\hat{a}' \\
\hat{c}'
\end{bmatrix}\hat{BS}_a &=& \begin{bmatrix} \eta_a^2 & i \sqrt{1-\eta_a^4} \\
i \sqrt{1-\eta_a^4} & \eta_a^2 \end{bmatrix}\begin{bmatrix}
\hat{a} \\
\hat{c}
\end{bmatrix} \\ &=& \begin{bmatrix}
\eta_a^2\hat{a} + i \sqrt{1-\eta_a^4} \hat{c} \\
i \sqrt{1-\eta_a^4} \hat{a} + \eta_a^2 \hat{c}
\end{bmatrix} , \nonumber
\end{eqnarray}
and a similar transformation is applied to the modes $b$ and $d$.

Phase shift transformation:
\begin{eqnarray}
    \hat{PS}_{ab}^\dagger \begin{bmatrix}
\hat{a}' \\
\hat{b}'
\end{bmatrix}\hat{PS}_{ab} = \begin{bmatrix} e^{i\phi} & 0 \\
0 & 1 \end{bmatrix}\begin{bmatrix}
\hat{a} \\
\hat{b}
\end{bmatrix} = \begin{bmatrix}
e^{i\phi}\hat{a} \\
\hat{b}
\end{bmatrix}. 
\end{eqnarray}

\subsection{Two mode squeezed coherent state input}
We consider as our input a TMSCS. It is created by passing two coherent sources with average photon numbers $\alpha^2$, $\beta^2$ and phases $\theta_1, \theta_2$ through a optical parametric amplifier with squeezing factor $r$ and squeezing angle $\xi$.
Thus, the input state of the interferometer $\ket{\Psi_{in}}_{ab}$ is given by:
\begin{eqnarray}
    \ket{\Psi_{in}}_{ab} = \hat{\mathbf{S}}_{a,b}(\mathbf{z}) \hat{\mathbf{D}}_b(\mathbf{b}) \hat{\mathbf{D}}_a(\mathbf{a})\ket{00}_{ab} ,
\label{eqn:TMSCS}
\end{eqnarray}
\\
where, $\hat{\mathbf{D}}_a(\mathbf{a}) = \exp(\mathbf{a} \hat{a}^\dagger - \mathbf{a}^* \hat{a})$ and $\hat{\mathbf{D}}_b(\mathbf{b}) = \exp(\mathbf{b} \hat{b}^\dagger - \mathbf{b}^* \hat{b})$ are the Displacement operators for modes $a$ and $b$ entering the interferometer with $\mathbf{a} = \alpha e^{i\theta_1}$ and $\mathbf{b} = \beta e^{i\theta_2}$. 

Further, the two mode squeezing operator implemented by the parametric amplifier is given by:
\begin{eqnarray}
    \mathbf{\hat{S}}_{a,b}(\mathbf{z}) &=& \exp(r(\hat{a}\hat{b}e^{-i\xi} - \hat{a}^\dagger \hat{b}^\dagger e^{i\xi})) ,
\end{eqnarray}
where, $\mathbf{z} = re^{i\xi}$.

Average photon number for TMSCS is given by:
\begin{eqnarray}
    \bar{n} = (\alpha^2 + \beta^2)\cosh(2r)-2\alpha\beta\sinh(2r)\cos(\Theta) + 2\sinh^2r \rm{,} \nonumber \\ 
    \label{eqn:pn_number_TMSCS}
\end{eqnarray}
where, $\Theta = \theta_1 + \theta_2 - \xi$. Clearly $\bar{n}$ is maximized for $\Theta = \pi$, regardless of the individual values of $\theta_1, \theta_2, \xi$.

\subsection{Phase sensitivity}
The relative phase of the two arms can be measured through its modulation of the photon number difference between the two output detectors of the interferometer. 
For brevity, we now drop the mode subscripts $a,b$ for the operators unless the context is not obvious. 
The photon number difference operator is defined as:
\begin{eqnarray}
    \hat{J}_3 = \hat{a}^{\dagger} \hat{a} - \hat{b}^{\dagger} \hat{b}.
\end{eqnarray}
The phase sensitivity of the interferometer is defined by the following relation:
\begin{eqnarray}
    \Delta \phi &=& \frac{\Delta \hat{J}_3}{\vert \frac{d\langle \hat{J}_3 \rangle}{d\phi} \vert} ,
\end{eqnarray}
where $\langle \hat{J}_3 \rangle = \langle \psi_{out}\vert \hat{J}_3 \vert \psi_{out}\rangle$ and $(\Delta \hat{J}_3)^2 = \langle \hat{J}_3^2 \rangle - \langle \hat{J_3} \rangle^2$.

With the above definitions the mean and variance of $\hat{J_3}$ can be evaluated as shown in Equations (\ref{eqn:J3mean},\ref{eqn:J3variance}). Since it is known that the phase sensitivity increases with increasing mean photon number of the source (TMSCS in this case), we work under the condition when $\bar{n}$ is maximum, i.e., $\Theta = \theta_1+\theta_2-\xi = \pi$. Within this constraint two of the three phases can be chosen independently. For simplicity, the expressions provided below assume $\theta_1=0$ and therefore $\theta_2=\pi+\xi$, where $\xi$ can now be varied freely. We have also assumed that the input coherent beams are strong, i.e., $\alpha^2,\beta^2 \gg 1$ and that the squeezing $e^{2r} \gg 1$. (The covariance matrix formalism is used to obtain more general expressions without these constraints. See Appendix \ref{Appendix_a} for details.) 

\begin{widetext}
\begin{eqnarray}
    \langle\hat{J_3}\rangle &=& \frac{1}{2}\eta_a\eta_b(\alpha+\beta)^2\cos(\xi)\sin(\phi)
    \label{eqn:J3mean}
\end{eqnarray}
\begin{eqnarray}
    (\Delta \hat{J_3})^2 &=& \frac{1}{4}\bigg{\{}2(\alpha^2+\beta^2)\eta_a^2\eta_b^2 + e^{-2r}(\alpha-\beta)^2\{(2\bar{\gamma_a}+1)\eta_a^2 + (2\bar{\gamma_b}+1)\eta_b^2 -2(\bar{\gamma_a}+\bar{\gamma_b}+1)\eta_a^2\eta_b^2\} \nonumber \\ 
    &+& e^{2r}(\alpha+\beta)^2\big{\{}\{(2\bar{\gamma_a}+1)\eta_a^2 + (2\bar{\gamma_b}+1)\eta_b^2 -2(\bar{\gamma_a}+\bar{\gamma_b}+1)\eta_a^2\eta_b^2\} \nonumber \\ 
    &-& \sin\xi\{(2\bar{\gamma_a}+1)\eta_a^2 + (2\bar{\gamma_b}+1)\eta_b^2 -2(\bar{\gamma_a}+\bar{\gamma_b}+1)\eta_a^2\eta_b^2\} \big{\}}\bigg{\}}.
    \label{eqn:J3variance}
\end{eqnarray}
\end{widetext}

It is clear from the above expressions that although the phase sensitivity is maximum for $\Theta = \pi$, the choice of the individual phases within that constraint affects the phase sensitivity without changing the mean photon number. Or in other words, by choosing the phases optimally the phase sensitivity can be further enhanced with same number of input photons. We show this in the following section.
\subsection{Enhancement in phase sensitivity}
\begin{figure*}[ht]  
\includegraphics[scale=0.4]{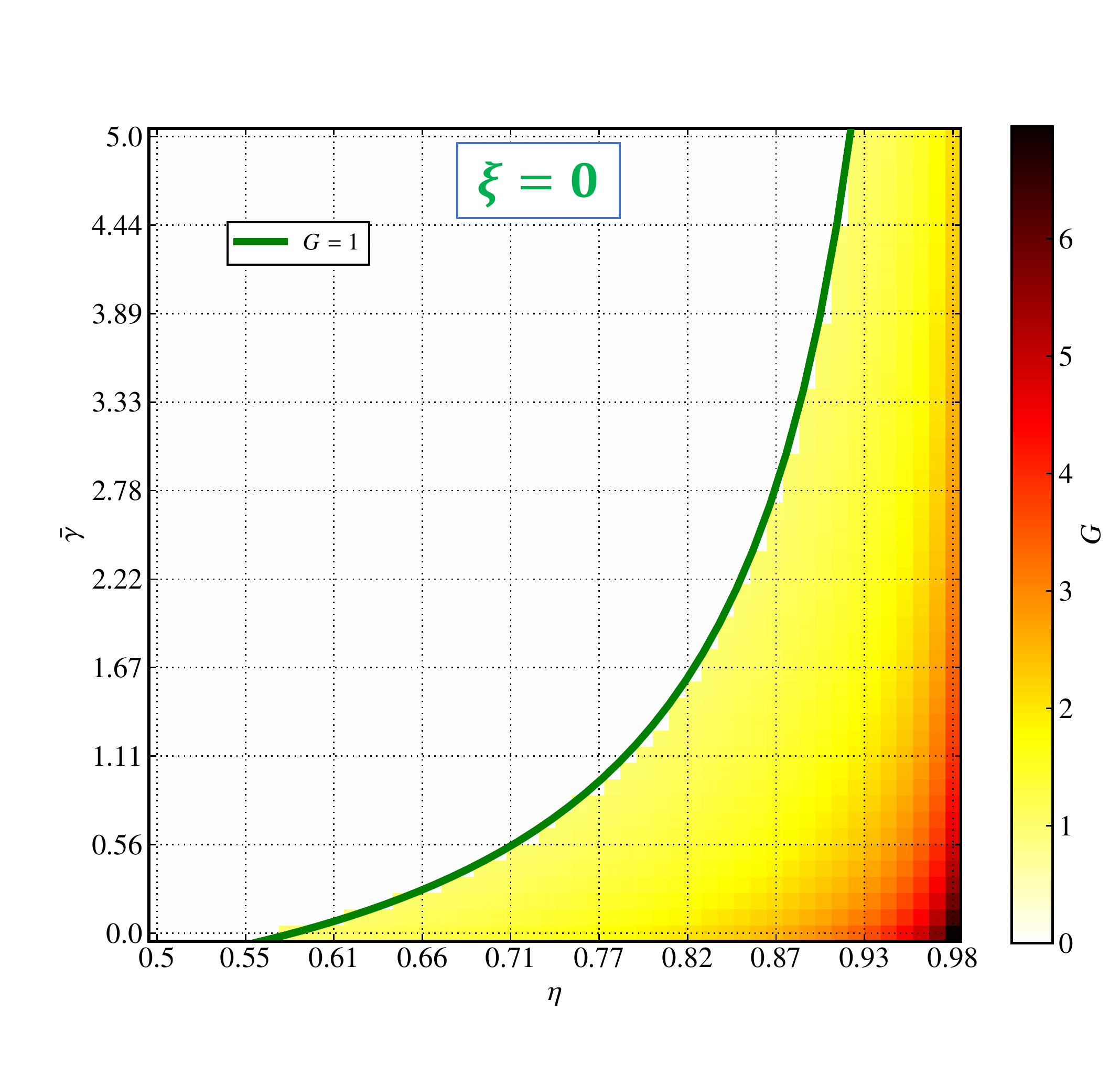}
\includegraphics[scale=0.4]{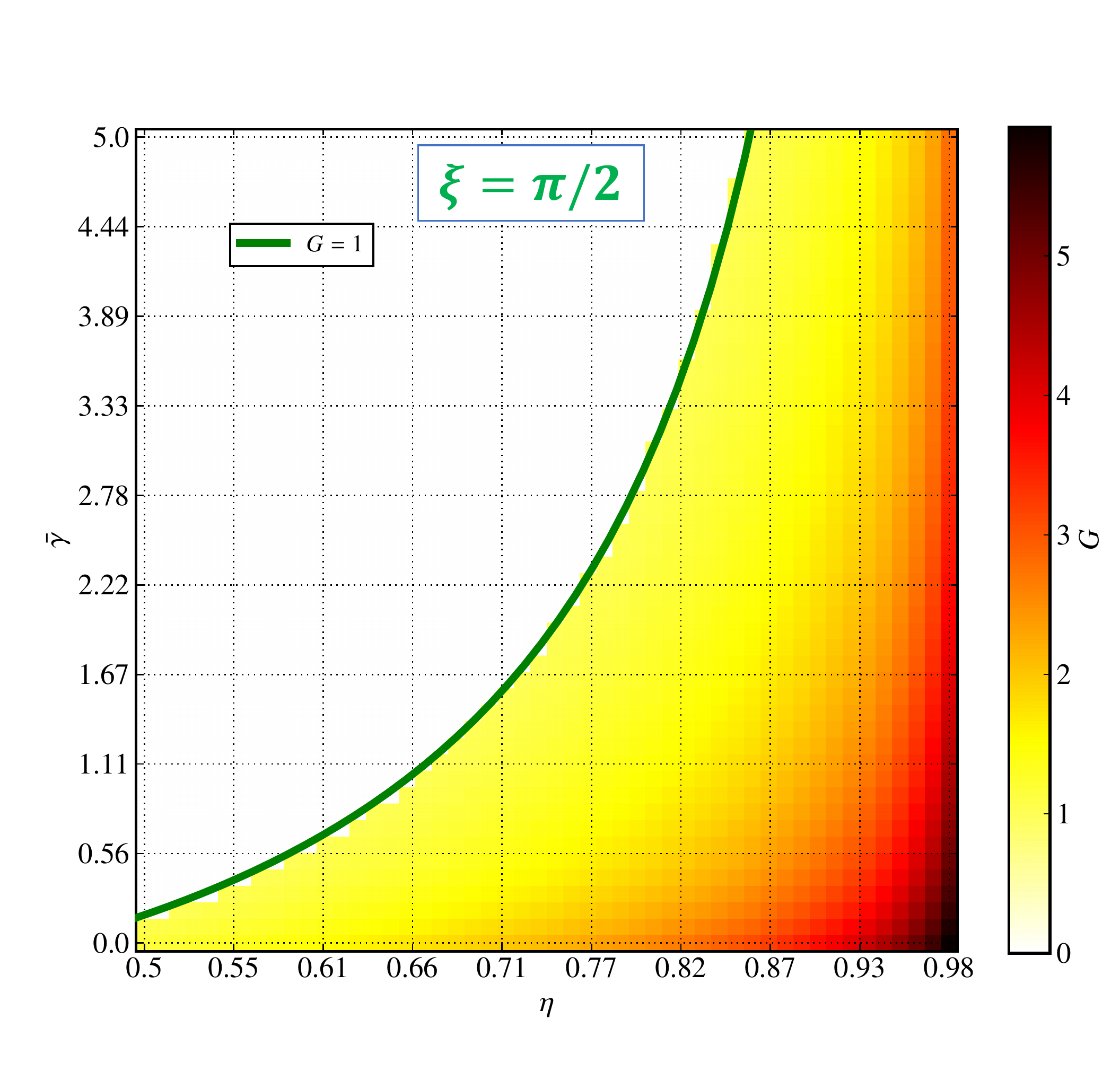}
\caption{\label{fig:one_arm} Gain in phase sensitivity of a MI  over the SN limit using a TMSCS input, assuming loss and noise in the upper arm only. Such a configuration is relevant for positioning and ranging applications where the upper arm acts as the target arm and the lower arm as a reference arm. In these plots, to indicate regions of sensitivity enhancement, we take $G=0$ whenever $G<1$. Loss is parameterized by $\eta$, where $\eta^2$ is the transmissivity of the upper arm, and noise is parameterized by $\bar{\gamma}$, i.e., the mean photon number of the thermal state used to model the noise. The G=1 line is shown in green which is also given by Eqn. \eqref{eqn:one_arm_G=1}. For all values of $\eta$ and $\bar{\gamma}$ below the green curve the SN limit can be broken. The gain factor $G$ can be increased by optimally choosing the squeezing angle. Also this can be done without changing the average photon number of the TMSCS and thus leads to an enhanced resistance to noise and loss without changing the input power. A larger region in the $\eta - \bar{\gamma}$ space gives $G>1$ for $\xi = \pi/2 - 0.05$ (Right), than for for $\xi=0$ (Left).}
\end{figure*}

The enhancement in phase sensitivity can be calculated by comparing the scaling of $\Delta \phi$ with mean photon number. If 
$(\Delta \phi)^2 = \frac{1}{G \bar{n}}$, then $G$ is defined as the phase sensitivity enhancement factor over the shot noise (SN) limit and $G=\frac{(\Delta \phi)^2}{(\Delta \phi_{SN})^2}$ \cite{xiao2020}. We will now look at this enhancement factor for the TMSCS input in different cases. Unless stated otherwise, the phase sensitivity is maximum at $\phi=0$ and the expression for $\Delta \phi$ will thus be evaluated at $\phi=0$.
\begin{itemize}
    \item Case 1: Phase sensitivity gain compared to the SN limit in the noiseless and loss-free case. If we set $\alpha=\beta$, and $\theta_1=\theta_2=\pi/2, \xi=0$ in Equation \eqref{eqn:pn_number_TMSCS}, the SN limit for the phase sensitivity in this case is given by:
    \begin{eqnarray}
        (\Delta \phi_{SN})^2 = \frac{1}{\bar{n}} = \frac{1}{2\vert\alpha\vert^2e^{2r}}.
    \end{eqnarray}
    On the other hand using Equations \eqref{eqn:J3mean} and \eqref{eqn:J3variance}, the same choice of input phases and setting $\eta_a=\eta_b=1$, we find the phase sensitivity in this case as:
    \begin{eqnarray}
    (\Delta \phi)^2 = \frac{1}{2\vert\alpha\vert^2e^{4r}} = \frac{1}{G \bar{n}},
    \label{eqn:pc_gain}
    \end{eqnarray}
    thus, the gain factor $G = e^{2r}$.
    \item Case 2: Loss in one arm of interferometer, i.e. $\eta_a = \eta$ and $\eta_b=1$. Also, the arms are noise-free, i.e., $\bar{\gamma_a}=\bar{\gamma_b}=0$ (except the vacuum fluctuations that appear because of $\hat{BS}_a(\eta_a^2)$). We assume $\theta_1=0$ and $\theta_2=\pi+\xi$, where $\xi$ can now be varied freely. Again $\alpha=\beta$. The phase sensitivity in this case is given by:
    \begin{eqnarray}
        (\Delta \phi)^2 = (\Delta \phi_{SN})^2\bigg( \frac{1}{e^{2r}\cos^2(\xi)} + \frac{(1-\eta^2)}{2\eta^2}\frac{(1-\sin\xi)}{\cos^2(\xi)}\bigg), \nonumber \\
        \label{eqn:our_opt}
    \end{eqnarray}
    for $\xi = 0$ this reduces to known result \cite{xiao2020}:
    \begin{eqnarray}
        (\Delta \phi)^2 = (\Delta \phi_{SN})^2\bigg( \frac{1}{e^{2r}} + \frac{(1-\eta^2)}{2\eta^2}\bigg).
        \label{eqn:xiao_opt}
    \end{eqnarray}
    To beat SN limit we need $G > 1$. In the latter case Eq.~\eqref{eqn:xiao_opt}, this translates to $\frac{(1-\eta^2)}{2\eta^2} < 1$, which is true for $\eta > 1/3$ \cite{xiao2020}. But, for the more general expression Eq.~\eqref{eqn:our_opt}, where the squeezing angle can be chosen freely, this limit for the tolerable loss can be further increased to $\eta > 1/5$. This is because the minimum value of $\frac{(1-\sin\xi)}{\cos^2(\xi)}$ is $1/2$ which happens when $\xi \to \pi/2$. But at the same time, at $\xi = \pi/2$ the first term in Eq.~\eqref{eqn:our_opt} blows up, killing the advantage. Nonetheless, it can be seen that for values of $\xi$ close to $\pi/2$, $\eta$ can be pushed below $1/3$ and can ultimately approach $1/5$. Larger the value of the squeezing parameter $r$, closer the optimum value for $\xi$ to $\pi/2$ and closer is the lower limit of $\eta$ to 1/5.

    \item Case 3: If we add thermal noise with mean photon number $\bar{\gamma_a} = \bar{\gamma}$, also in the upper arm then the phase sensitivity becomes:
    \begin{eqnarray}
        (\Delta \phi)^2 = (\Delta \phi_{SN})^2&\bigg(& \frac{1}{e^{2r}\cos^2(\xi)} + \nonumber \\
         &(&2\bar{\gamma}+1)\frac{(1-\eta^2)}{2\eta^2}\frac{1-\sin(\xi)}{\cos^2(\xi)}\bigg) \rm{,} \nonumber \\
    \end{eqnarray}
    again, the maximum noise and loss that can be tolerated is increased by the choice $\xi \to \pi/2$. It is given by the condition (also see Fig. \ref{fig:one_arm}):
    \begin{eqnarray}
    \label{eqn:one_arm_G=1}
        (2\bar{\gamma}+1)\frac{(1-\eta^2)}{4\eta^2} \leq 1.
    \end{eqnarray}

    \item Case 4: Equal loss and noise in the two arms. The expression for phase sensitivity in this case becomes (choice of phases as in the previous cases):
    \begin{eqnarray}
        (\Delta \phi)^2 = \frac{(\Delta \phi_{SN})^2} {\cos^2(\xi)}\bigg( \frac{1}{e^{2r}} + (2\bar{\gamma}+1)\frac{(1-\eta^2)}{\eta^2}\bigg).
    \end{eqnarray}
    The gain in this case is maximum for $\xi = 0$ and is given by:
    \begin{eqnarray}
        G = \bigg(\frac{1}{e^{2r}} + (2\bar{\gamma}+1)\frac{(1-\eta^2)}{\eta^2}\bigg)^{-1},
    \end{eqnarray}
    with, $G>1$, whenever the following condition is satisfied:
    \begin{eqnarray}
        (2\bar{\gamma}+1)\frac{(1-\eta^2)}{\eta^2} \leq 1.
        \label{eqn:green_line_sym}
    \end{eqnarray}
\end{itemize}

\begin{figure}[h]  
\includegraphics[scale=0.38]{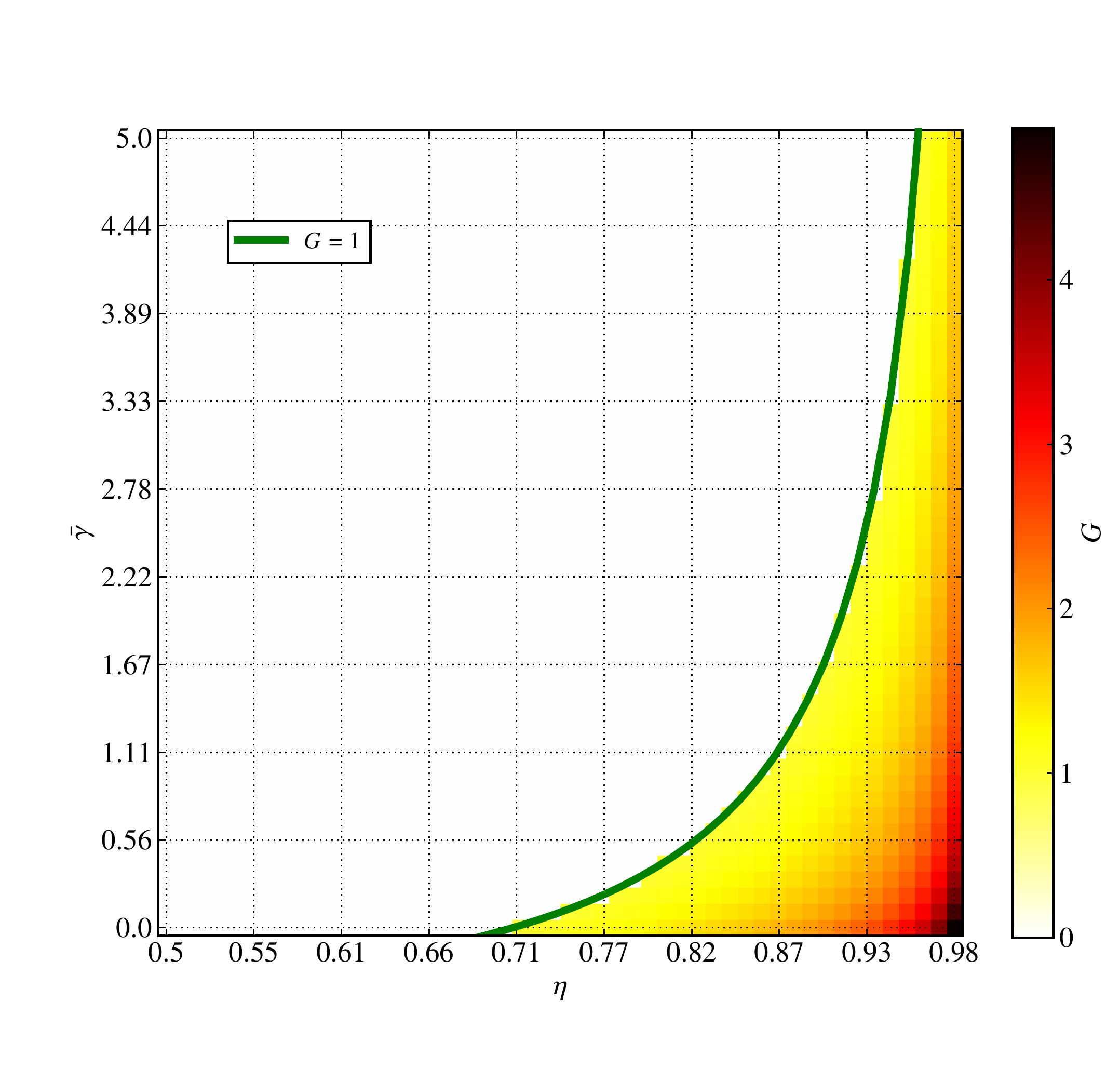}
\caption{\label{fig:both_arm} Gain in phase sensitivity of a MI over the SN limit using a TMSCS input, assuming same loss and noise in both arms. Such a symmetric configuration is relevant for applications like gyroscopes \cite{xiao2020}. In these plots, to indicate regions of sensitivity enhancement, we take $G=0$ whenever $G<1$. Loss is parameterized by $\eta$, where $\eta^2$ is the transmissivity of each arm, and noise is parameterized by $\bar{\gamma}$, i.e., the mean photon number of the thermal state used to model the noise. The G=1 line is shown in green which is also given by Eq. \eqref{eqn:green_line_sym}. For all values of $\eta$ and $\bar{\gamma}$ below the curve the SN limit can be broken. The gain factor $G$ is maximum for choice of squeezing angle $\xi=0$.}
\end{figure}

\section{Radiation pressure error}
\label{sec:radiation_pressure}
For applications of the MI such as gravitational wave detectors, where the input power is high, and the end mirrors by design are free to move, radiation pressure fluctuation is a major source of deterioration in phase sensitivity. It was shown by Caves \cite{caves1981}, that using a single-mode squeezed vacuum state as one of the interferometer inputs (coherent state as the other input) can reduce the photon counting error, i.e., fluctuations in photon number difference by a factor of $e^r$. This is on account of reduced uncertainty in one of the quadratures of the squeezed state compared to vacuum and coherent state. At the same time, there is an increase in radiation pressure error by the same factor of $e^r$, due to increased uncertainty in the other quadrature.

Here we will show that the same kind of trade-off between photon counting error and radiation pressure error exists for two-mode entangled squeezed coherent states. Now, consider a TMSCS entering the two input ports of an MI. Let the two modes be represented by $\mathbf{a}$ and $\mathbf{b}$. After interfering at the 50:50 beam-splitter once, let's say that the output modes are $\mathbf{a'}$ and $\mathbf{b'}$, which are given by Eq. \eqref{eqn:bs}.
The error in measurement of path difference $\Delta z$ is then given by the difference in radiation pressures, $\Delta \mathcal{P}$, applied by these modes on the two end mirrors of the MI. The radiation pressure is in turn proportional to the intensity or incident photon numbers. Therefore, for a measurement duration $\tau$, and end mirrors each having mass $m$, the error in $z$ due to quantum mechanical fluctuations in radiation pressure is given by:
\begin{eqnarray}
\Delta z_{rp} = \frac{\Delta\mathcal{P} \tau}{2m}
\label{eqn:rperror}
\end{eqnarray}
which is equivalent to a relative phase measurement error of:
\begin{eqnarray}
    \Delta \phi_{rp} = \frac{\omega}{c}\frac{\Delta\mathcal{P} \tau}{2m}
    \label{eqn:phase_rp_error}
\end{eqnarray} 
where $\omega$ is the angular frequency of the source and $c$ is the speed of light,
\begin{eqnarray}
\mathcal{P} = \frac{2\hbar \omega} {c} (\hat{a}'^\dagger \hat{a}' - \hat{b}'^\dagger \hat{b}')
\label{eqn:momentum}
\end{eqnarray}
In terms of the original input modes $\hat{a}$ and $\hat{b}$ in Eq. \eqref{eqn:bs}, Eq. \eqref{eqn:momentum}, reads:
\begin{eqnarray}
\mathcal{P} = \frac{2\hbar \omega} {c} (\hat{a}^\dagger \hat{b} - \hat{b}^\dagger \hat{a})
\label{eqn:momentum_inp}
\end{eqnarray}
The error in the momentum difference due to quantum fluctuations can be quantified by the variance in $\mathcal{P}$ and is given by:
\begin{eqnarray}
(\Delta \mathcal{P})^2 = \langle\mathcal{P}^2\rangle - \langle \mathcal{P} \rangle^2
\label{eqn:radvar}
\end{eqnarray}
where expectation values are taken over the TMSCS given by Eq. \eqref{eqn:TMSCS},
therefore,
\begin{eqnarray}
(\Delta &&\mathcal{P})^2 = (\alpha^2 + \beta^2)\cosh(4r) - \nonumber\\ && 2\alpha \beta \sinh(4r)\cos(\theta_1 + \theta_2 - \xi) + \frac{\sinh^2(2r)}{2}
\label{eqn:Pvariance}
\end{eqnarray}
Putting $\mathbf{a} = \mathbf{b}$ in Eq. \eqref{eqn:Pvariance} and then substituting it in Eq. \eqref{eqn:phase_rp_error} leads to a simple expression for the radiation pressure error:
\begin{eqnarray}
\Delta \phi_{rp} &&= \frac{\hbar \omega \tau}{mc} \Big[2\alpha^2 \cosh(4r) - 2\alpha^2 \sinh(4r)\cos(2\theta-\xi) \nonumber \\ &&+ \frac{\sinh^2(2r)}{2} \Big]^{\frac{1}{2}}
\end{eqnarray}

Now, let us consider the condition for which the photon counting error is minimum. This happens whenever the average photon number of the TMSCS source is maximum. In terms of choice of input phases, this translates to the condition $\Theta = \pi$, i.e., in this case $2\theta-\xi = \pi$. Clearly under such an assumption, the radiation pressure error given by Eq. \eqref{eqn:rperror}, is maximum for a given value of $\alpha$. 
Further, the equivalent of Eq. \eqref{eqn:pc_gain} is given by:
\begin{eqnarray}
    (\Delta \phi)^2 = {2\vert\alpha\vert^2e^{4r}}
    \label{eqn:rp_gain}
\end{eqnarray}
Therefore, the phase error is increased, by a gain factor $e^{2r}$ compared to the case with no squeezing, which exactly compensates the reduction in phase error due photon counting fluctuations.

On the other hand if the radiation pressure error has to be minimized then the average photon number must be minimized. For given $\alpha$, this leads to the following condition on the choice of phases: $\Theta = 0$, i.e, $2\theta - \xi = 0$.
The phase error due to radiation pressure in this case reduces by a factor of $e^{2r}$. At the same time, for such a choice of the phases, the photon counting error is maximum and the corresponding negative gain factor is $G=e^{-2r}$. Therefore, there exists a trade-off between photon counting error and radiation pressure error for the TMSCS input for a MI, just like single mode squeezed vacuum state inputs \cite{caves1981}.
\section{Conclusions}
\label{sec:conclusions}
In this paper we calculated the phase sensitivity of a Michelson interferometer for a two mode squeezed coherent state input. Squeezed states, by utilizing reduced quantum fluctuations in one mode, can enhance the phase sensitivity compared to coherent states. This enhancement increases with increasing squeezing factor $r$. Of course this is at the cost of higher input pump power. This enhancement is tolerant to some degree of loss in the interferometer arms. It was shown in previous work that in the case with equal loss in both arms, a loss of up to 50\% (3 dB) can be tolerated while maintaining the advantages of squeezed states. For the asymmetric case with loss only in one arm, this limit was shown to be 66.7\%. In this work we show that this limit can be further increased to 80\% by appropriately choosing the squeezing angle $\xi$. We show that $\xi \to \pi/2$, is the optimal choice in this case. For the symmetric case, $\xi = 0$ was found to be optimal. Further, we also include the effects of thermal noise on the phase sensitivity. We find that sensitivity enhancement can be achieved even in the noisy case and the choice of optimal squeezing angles remain the same. It is important to note that phase sensitivity enhancement for a given squeezing factor $r$ is calculated compared to a coherent source with the same number of input photons, the so called shot noise limit. In similar spirit, we show here that by choosing the appropriate squeezing angle, while keeping the total number of photons going into the interferometer fixed, we can enhance the phase sensitivity. In this way we propose an alternative mechanism to improve the sensitivity of an interferometer, that is, by optimally choosing the input phases like the squeezing angle or coherent source phases. Finally, we also show that there exists a trade-off between photon-counting error usually relevant for low power applications and the radiation pressure error relevant for high power uses. One can be reduced at the expense of the other, with the reduction and increase scaling like $e^{-2r}$ and $e^{2r}$ respectively for the TMSCS input, thus maintaining the standard quantum limit. 
\begin{acknowledgments}
We wish to acknowledge the support of the Army Research Office and the Air Force Office of Scientific Research. PB and HL also acknowledge the support of the Binational Science Foundation.

We are honored to dedicate this paper to the memory of Jonathan P. Dowling. Jonathan was an early advocate of the emergent quantum technology from its early days. He was a great mentor, colleague, and friend.  We will miss him for a long long time.
\end{acknowledgments}

\section{Covariance matrix formalism}
\label{Appendix_a}
It is convenient to work with the covariance matrix formalism since we deal with states and interferometer operations which are Gaussian. A N-mode Gaussian state has Wigner function of following form \cite{weedbrook2012gaussian}
\begin{equation}
    W(x) = \frac{\exp{\left[-\frac{1}{2}(x-\left\langle\hat{x}\right\rangle)^T \sigma^{-1}(x-\left\langle\hat{x}\right\rangle)\right]}}{(2\pi)^N \sqrt{det \sigma}}
\end{equation}
which is completely determined by: a $2N$-dimensional mean vector $\hat{x} = (\hat{q}_1,\hat{p}_1,...\hat{q}_N,\hat{p}_N)^T$ and $2N\times2N$ covariance matrix
\begin{equation}
    \sigma_{jk} = \frac{1}{2}\left\langle \left\{ (\hat{x}_j-\langle\hat{x}_j\rangle),(\hat{x}_k-\langle\hat{x}_k\rangle) \right\} \right\rangle,
\end{equation}
where \{*,*\} denotes anticommutator, $\hat{q}_k = \frac{1}{\sqrt2}(\hat{a}^\dagger_k+\hat{a}_k)$ and $\hat{p}_k=\frac{1}{\sqrt2i}(\hat{a}^\dagger_k-\hat{a}_k)$ are the quadrature operators associated with the k${th}$ mode defined via standard creation $(\hat{a}^\dagger_k)$ and annihilation $(\hat{a}_k)$ operators.
Action of any Gaussian unitary can be represented by a symplectic matrix $S$ with the transformations
\begin{equation}
\hat{x} \longrightarrow  S\hat{x}, \hspace{1cm} \sigma \longrightarrow S \sigma S^T.
\end{equation} 
$\hat{x}$ and $\sigma$ at input are evolved using symplectic matrices of two-mode squeezing unitary $(S_{TMS})$, beamsplitter $(S_{BS})$ and phase shift $(S_{PS})$ \cite{gard2016advances} to get their final form at the output ports. 

For the purpose of calculating desired expectation values we construct the characteristic function, still a Gaussian, from final $\langle\hat{x}\rangle$ and $\sigma$ matrices as
\begin{equation}
\chi(\Lambda) = \exp\left\{ -\frac{1}{2}\Lambda^T\sigma\Lambda +i\Lambda^T\left\langle \hat{x} \right\rangle\right\},
\end{equation}
with $\Lambda = (\Lambda_{q_1},\Lambda_{p_1},\Lambda_{q_2},\Lambda_{p_2})^T$ for the two mode case considered in this work. Expectation values of position and momentum quadrature operators for the k$th$ mode are calculated by using standard expressions
\begin{equation} 
\expval{q_k^m} = \left( -i \right)^m \left.\frac{\partial^m\chi}{\partial\Lambda^m_{q_k}}\right\vert_{\Lambda=0} \textrm{and }
\expval{p_k^m} = \left( -i \right)^m \left.\frac{\partial^m\chi}{\partial\Lambda^m_{p_k}}\right\vert_{\Lambda=0},
\end{equation}
which are used for intensity and variance calculation as shown in \cite{gard2017nearly}.
Using the formalism discussed above we find the general expressions for mean and variance of the photon number difference operator $\hat{J}_3$. These are given as:
\begin{widetext}
\begin{multline*}
\langle \hat{J}_3 \rangle = \frac{1}{4} e^{-2 r} (\alpha ^2\eta_b^2-\alpha ^2\eta_a^2+2 \alpha  \beta \eta_b^2 \sin (\theta _1-\theta _2)+2 \alpha  \beta \eta_a^2 \sin (\theta _1-\theta _2)+\beta ^2\eta_b^2 \sin (\xi - 2 \theta _2)+ \\
\beta ^2\eta_b^2+ \beta ^2\eta_a^2 \sin (\xi -2 \theta _2)-\beta ^2\eta_a^2+\gamma ^2-\eta ^2-4\eta_b^2\bar{\gamma}_a e^{2 r}+4\bar{\gamma}_a e^{2 r}+4\eta_a^2\bar{\gamma}_b e^{2 r}-4\bar{\gamma}_b e^{2 r}+ \\
\alpha ^2 (e^{4 r}-1) (\gamma ^2+\eta ^2) \sin (\xi -2 \theta _1)+\alpha ^2\eta_b^2 e^{4 r}-\alpha ^2\eta_a^2 e^{4 r}-2 \alpha  \beta  (e^{4 r}-1) (\gamma ^2-\eta ^2) \cos (-\theta _1-\theta _2+\xi )+ \\
2 \alpha  \beta \eta_b^2 e^{4 r} \sin (\theta _1-\theta _2)+
2 \alpha  \beta \eta_a^2 e^{4 r} \sin (\theta _1-\theta _2)-\beta ^2\eta_b^2 e^{4 r} \sin (\xi -2 \theta _2)+\beta ^2\eta_b^2 e^{4 r}-\beta ^2\eta_a^2 e^{4 r} \sin (\xi -2 \theta _2)- \\
\beta ^2\eta_a^2 e^{4 r}-2\eta_b^2 e^{2 r}+\gamma ^2 e^{4 r}+2\eta_a^2 e^{2 r}-\eta ^2 e^{4 r})
\end{multline*}

\begin{multline*}
(\Delta \hat{J}_3)^2 = 
\frac{1}{8} e^{-4 r} (2 e^{2 r}\eta_b^2-4 e^{4 r}\eta_b^2+2 e^{6 r}\eta_b^2+2 e^{2 r} \alpha ^2\eta_b^2+2 e^{6 r} \alpha ^2\eta_b^2+4 e^{2 r} \bar{\gamma}_b \alpha ^2\eta_b^2+4 e^{6 r} \bar{\gamma}_b \alpha ^2\eta_b^2+2 e^{2 r} \beta ^2\eta_b^2+ \\
2 e^{6 r} \beta ^2\eta_b^2+ 4 e^{2 r} \bar{\gamma}_b \beta ^2\eta_b^2+4 e^{6 r} \bar{\gamma}_b \beta ^2\eta_b^2- 4 e^{2 r}\eta_a^2\eta_b^2+6 e^{4 r}\eta_a^2\eta_b^2-4 e^{6 r}\eta_a^2\eta_b^2+e^{8 r}\eta_a^2\eta_b^2-4 e^{2 r} \alpha ^2\eta_a^2\eta_b^2+4 e^{4 r} \alpha ^2\eta_a^2\eta_b^2- \\
4 e^{6 r} \alpha ^2\eta_a^2\eta_b^2+2 e^{8 r} \alpha ^2\eta_a^2\eta_b^2-4 e^{2 r} \bar{\gamma}_a \alpha ^2\eta_a^2\eta_b^2-4 e^{6 r} \bar{\gamma}_a \alpha ^2\eta_a^2\eta_b^2-4 e^{2 r} \bar{\gamma}_b \alpha ^2\eta_a^2\eta_b^2-4 e^{6 r} \bar{\gamma}_b \alpha ^2\eta_a^2\eta_b^2+ \\
2 \alpha ^2\eta_a^2\eta_b^2-4 e^{2 r} \beta ^2\eta_a^2\eta_b^2+4 e^{4 r} \beta ^2\eta_a^2\eta_b^2-4 e^{6 r} \beta ^2\eta_a^2\eta_b^2+2 e^{8 r} \beta ^2\eta_a^2\eta_b^2-4 e^{2 r} \bar{\gamma}_a \beta ^2\eta_a^2\eta_b^2-4 e^{6 r} \bar{\gamma}_a \beta ^2\eta_a^2\eta_b^2-\\
4 e^{2 r} \bar{\gamma}_b \beta ^2\eta_a^2\eta_b^2-4 e^{6 r} \bar{\gamma}_b \beta ^2\eta_a^2\eta_b^2+2 \beta ^2\eta_a^2 \eta_b^2-4 e^{2 r} \bar{\gamma}_a\eta_a^2\eta_b^2+8 e^{4 r} \bar{\gamma}_a\eta_a^2\eta_b^2-4 e^{6 r} \bar{\gamma}_a \eta_a^2\eta_b^2- \\
4 e^{2 r} \bar{\gamma}_b\eta_a^2\eta_b^2+8 e^{4 r} \bar{\gamma}_b\eta_a^2\eta_b^2-4 e^{6 r} \bar{\gamma}_b\eta_a^2\eta_b^2+16 e^{4 r} \bar{\gamma}_a \bar{\gamma}_b\eta_a^2\eta_b^2+\eta ^2\eta_b^2-8 e^{4 r} \bar{\gamma}_a\eta_b^2+4 e^{2 r} \bar{\gamma}_b\eta_b^2- \\
8 e^{4 r} \bar{\gamma}_b\eta_b^2+4 e^{6 r} \bar{\gamma}_b\eta_b^2-16 e^{4 r} \bar{\gamma}_a \bar{\gamma}_b\eta_b^2-(-1+e^{4 r})^2 (2 \alpha ^2+2 \beta ^2+1)\eta_a^2 \cos (2 \phi )\eta_b^2+2 e^{8 r} \alpha  \beta \eta_a^2 \cos (\xi +2 \phi -\theta _1-\theta _2)\eta_b^2- \\
2 \alpha  \beta \eta_a^2 \cos (\xi +2 \phi -\theta _1-\theta _2)\eta_b^2+2 e^{8 r} \alpha  \beta \eta_a^2 \cos (\xi -2 \phi -\theta _1-\theta _2)\eta_b^2-2 \alpha  \beta \eta_a^2 \cos (\xi -2 \phi -\theta _1-\theta _2)\eta_b^2- \\
2 e^{2 r} \alpha ^2 \sin (\xi -2 \theta _1)\eta_b^2+2 e^{6 r} \alpha ^2 \sin (\xi -2 \theta _1)\eta_b^2-4 e^{2 r} \bar{\gamma}_b \alpha ^2 \sin (\xi -2 \theta _1)\eta_b^2+4 e^{6 r} \bar{\gamma}_b \alpha ^2 \sin (\xi -2 \theta _1)\eta_b^2- \\
4 e^{2 r} \bar{\gamma}_a \alpha ^2\eta_a^2 \sin (\xi -2 \theta _1)\eta_b^2+4 e^{6 r} \bar{\gamma}_a \alpha ^2\eta_a^2 \sin (\xi -2 \theta _1)\eta_b^2+4 e^{2 r} \bar{\gamma}_b \alpha ^2\eta_a^2 \sin (\xi -2 \theta _1)\eta_b^2- \\
4 e^{6 r} \bar{\gamma}_b \alpha ^2\eta_a^2 \sin (\xi -2 \theta _1)\eta_b^2+2 e^{2 r} \alpha ^2\eta_a^2 \sin (\xi +2 \phi -2 \theta _1)\eta_b^2-2 e^{6 r} \alpha ^2\eta_a^2 \sin (\xi +2 \phi -2 \theta _1)\eta_b^2- \\
2 e^{2 r} \alpha ^2\eta_a^2 \sin (\xi -2 \phi -2 \theta _1)\eta_b^2+2 e^{6 r} \alpha ^2\eta_a^2 \sin (\xi -2 \phi -2 \theta _1)\eta_b^2+8 e^{2 r} \bar{\gamma}_a \alpha  \beta \eta_a^2 \sin (\theta _1-\theta _2)\eta_b^2+ \\
8 e^{6 r} \bar{\gamma}_a \alpha  \beta \eta_a^2 \sin (\theta _1-\theta _2)\eta_b^2-8 e^{2 r} \bar{\gamma}_b \alpha  \beta \eta_a^2 \sin (\theta _1-\theta _2)\eta_b^2-8 e^{6 r} \bar{\gamma}_b \alpha  \beta \eta_a^2 \sin (\theta _1-\theta _2)\eta_b^2+ \\
4 e^{2 r} \alpha  \beta  \sin (\theta _1-\theta _2)\eta_b^2+4 e^{6 r} \alpha  \beta  \sin (\theta _1-\theta _2)\eta_b^2+8 e^{2 r} \bar{\gamma}_b \alpha  \beta  \sin (\theta _1-\theta _2)\eta_b^2+ \\
8 e^{6 r} \bar{\gamma}_b \alpha  \beta  \sin (\theta _1-\theta _2)\eta_b^2+2 e^{2 r} \beta ^2 \sin (\xi -2 \theta _2)\eta_b^2-2 e^{6 r} \beta ^2 \sin (\xi -2 \theta _2)\eta_b^2+4 e^{2 r} \bar{\gamma}_b \beta ^2 \sin (\xi -2 \theta _2)\eta_b^2- \\
4 e^{6 r} \bar{\gamma}_b \beta ^2 \sin (\xi -2 \theta _2)\eta_b^2+4 e^{2 r} \bar{\gamma}_a \beta ^2\eta_a^2 \sin (\xi -2 \theta _2)\eta_b^2-4 e^{6 r} \bar{\gamma}_a \beta ^2\eta_a^2 \sin (\xi -2 \theta _2)\eta_b^2- \\
4 e^{2 r} \bar{\gamma}_b \beta ^2\eta_a^2 \sin (\xi -2 \theta _2)\eta_b^2+4 e^{6 r} \bar{\gamma}_b \beta ^2\eta_a^2 \sin (\xi -2 \theta _2)\eta_b^2-2 e^{2 r} \beta ^2\eta_a^2 \sin (\xi +2 \phi -2 \theta _2)\eta_b^2+2 e^{6 r} \beta ^2\eta_a^2 \sin (\xi +2 \phi -2 \theta _2)\eta_b^2+ \\
2 e^{2 r} \beta ^2\eta_a^2 \sin (\xi -2 \phi -2 \theta _2)\eta_b^2-2 e^{6 r} \beta ^2\eta_a^2 \sin (\xi -2 \phi -2 \theta _2)\eta_b^2+2 e^{2 r}\eta_a^2-4 e^{4 r}\eta_a^2+2 e^{6 r}\eta_a^2+2 e^{2 r} \alpha ^2\eta_a^2+2 e^{6 r} \alpha ^2\eta_a^2+ \\
4 e^{2 r} \bar{\gamma}_a \alpha ^2\eta_a^2+4 e^{6 r} \bar{\gamma}_a \alpha ^2\eta_a^2+2 e^{2 r} \beta ^2\eta_a^2+2 e^{6 r} \beta ^2\eta_a^2+4 e^{2 r} \bar{\gamma}_a \beta ^2\eta_a^2+4 e^{6 r} \bar{\gamma}_a \beta ^2\eta_a^2+4 e^{2 r} \bar{\gamma}_a\eta_a^2- \\
8 e^{4 r} \bar{\gamma}_a\eta_a^2+4 e^{6 r} \bar{\gamma}_a\eta_a^2-8 e^{4 r} \bar{\gamma}_b\eta_a^2-16 e^{4 r} \bar{\gamma}_a \bar{\gamma}_b\eta_a^2+8 e^{4 r} \bar{\gamma}_a+8 e^{4 r} \bar{\gamma}_b+16 e^{4 r} \bar{\gamma}_a \bar{\gamma}_b-4 (-1+e^{4 r}) \alpha  \beta  (e^{4 r}\eta_b^2\eta_a^2+ \\
\gamma ^2\eta_a^2+e^{2 r} ((-2 (\bar{\gamma}_a+1)\eta_a^2-2 \bar{\gamma}_b (\eta ^2-1)+1)\eta_b^2+(2 \bar{\gamma}_a+1)\eta_a^2)) \cos (\xi -\theta _1-\theta _2)+2 e^{2 r} \alpha ^2\eta_a^2 \sin (\xi -2 \theta _1)- \\
2 e^{6 r} \alpha ^2\eta_a^2 \sin (\xi -2 \theta _1)+4 e^{2 r} \bar{\gamma}_a \alpha ^2\eta_a^2 \sin (\xi -2 \theta _1)-4 e^{6 r} \bar{\gamma}_a \alpha ^2\eta_a^2 \sin (\xi -2 \theta _1)-4 e^{2 r} \alpha  \beta \eta_a^2 \sin (\theta _1-\theta _2)- \\
4 e^{6 r} \alpha  \beta \eta_a^2 \sin (\theta _1-\theta _2)-8 e^{2 r} \bar{\gamma}_a \alpha  \beta \eta_a^2 \sin (\theta _1-\theta _2)-8 e^{6 r} \bar{\gamma}_a \alpha  \beta \eta_a^2 \sin (\theta _1-\theta _2)-2 e^{2 r} \beta ^2 \eta_a ^2 \sin (\xi -2 \theta _2)+ \\
2 e^{6 r} \beta ^2 \eta_a^2 \sin (\xi -2 \theta _2)-4 e^{2 r} \bar{\gamma}_a \beta ^2\eta_a^2 \sin (\xi -2 \theta _2)+4 e^{6 r} \bar{\gamma}_a \beta ^2\eta_a^2 \sin (\xi -2 \theta_2))
\end{multline*}  

\end{widetext}
\bibliography{refs}

\begin{thebibliography}{33}%
\makeatletter
\providecommand \@ifxundefined [1]{%
 \@ifx{#1\undefined}
}%
\providecommand \@ifnum [1]{%
 \ifnum #1\expandafter \@firstoftwo
 \else \expandafter \@secondoftwo
 \fi
}%
\providecommand \@ifx [1]{%
 \ifx #1\expandafter \@firstoftwo
 \else \expandafter \@secondoftwo
 \fi
}%
\providecommand \natexlab [1]{#1}%
\providecommand \enquote  [1]{``#1''}%
\providecommand \bibnamefont  [1]{#1}%
\providecommand \bibfnamefont [1]{#1}%
\providecommand \citenamefont [1]{#1}%
\providecommand \href@noop [0]{\@secondoftwo}%
\providecommand \href [0]{\begingroup \@sanitize@url \@href}%
\providecommand \@href[1]{\@@startlink{#1}\@@href}%
\providecommand \@@href[1]{\endgroup#1\@@endlink}%
\providecommand \@sanitize@url [0]{\catcode `\\12\catcode `\$12\catcode
  `\&12\catcode `\#12\catcode `\^12\catcode `\_12\catcode `\%12\relax}%
\providecommand \@@startlink[1]{}%
\providecommand \@@endlink[0]{}%
\providecommand \url  [0]{\begingroup\@sanitize@url \@url }%
\providecommand \@url [1]{\endgroup\@href {#1}{\urlprefix }}%
\providecommand \urlprefix  [0]{URL }%
\providecommand \Eprint [0]{\href }%
\providecommand \doibase [0]{http://dx.doi.org/}%
\providecommand \selectlanguage [0]{\@gobble}%
\providecommand \bibinfo  [0]{\@secondoftwo}%
\providecommand \bibfield  [0]{\@secondoftwo}%
\providecommand \translation [1]{[#1]}%
\providecommand \BibitemOpen [0]{}%
\providecommand \bibitemStop [0]{}%
\providecommand \bibitemNoStop [0]{.\EOS\space}%
\providecommand \EOS [0]{\spacefactor3000\relax}%
\providecommand \BibitemShut  [1]{\csname bibitem#1\endcsname}%
\let\auto@bib@innerbib\@empty
\bibitem [{\citenamefont {Zwierz}\ \emph {et~al.}(2012)\citenamefont {Zwierz},
  \citenamefont {P{\'{e}}rez-Delgado},\ and\ \citenamefont {Kok}}]{kok2012}%
  \BibitemOpen
  \bibfield  {author} {\bibinfo {author} {\bibfnamefont {Marcin}\ \bibnamefont
  {Zwierz}}, \bibinfo {author} {\bibfnamefont {Carlos~A.}\ \bibnamefont
  {P{\'{e}}rez-Delgado}}, \ and\ \bibinfo {author} {\bibfnamefont {Pieter}\
  \bibnamefont {Kok}},\ }\bibfield  {title} {\enquote {\bibinfo {title}
  {Ultimate limits to quantum metrology and the meaning of the heisenberg
  limit},}\ }\href {\doibase 10.1103/physreva.85.042112} {\bibfield  {journal}
  {\bibinfo  {journal} {Physical Review A}\ }\textbf {\bibinfo {volume} {85}}
  (\bibinfo {year} {2012}),\ 10.1103/physreva.85.042112}\BibitemShut {NoStop}%
\bibitem [{\citenamefont {Caves}(1981)}]{caves1981}%
  \BibitemOpen
  \bibfield  {author} {\bibinfo {author} {\bibfnamefont {Carlton~M.}\
  \bibnamefont {Caves}},\ }\bibfield  {title} {\enquote {\bibinfo {title}
  {Quantum-mechanical noise in an interferometer},}\ }\href {\doibase
  10.1103/PhysRevD.23.1693} {\bibfield  {journal} {\bibinfo  {journal} {Phys.
  Rev. D}\ }\textbf {\bibinfo {volume} {23}},\ \bibinfo {pages} {1693--1708}
  (\bibinfo {year} {1981})}\BibitemShut {NoStop}%
\bibitem [{\citenamefont {Pezz{\'{e}}}\ and\ \citenamefont
  {Smerzi}(2008)}]{pezze2008}%
  \BibitemOpen
  \bibfield  {author} {\bibinfo {author} {\bibfnamefont {Luca}\ \bibnamefont
  {Pezz{\'{e}}}}\ and\ \bibinfo {author} {\bibfnamefont {Augusto}\ \bibnamefont
  {Smerzi}},\ }\bibfield  {title} {\enquote {\bibinfo {title} {Mach-zehnder
  interferometry at the heisenberg limit with coherent and squeezed-vacuum
  light},}\ }\href {\doibase 10.1103/physrevlett.100.073601} {\bibfield
  {journal} {\bibinfo  {journal} {Physical Review Letters}\ }\textbf {\bibinfo
  {volume} {100}} (\bibinfo {year} {2008}),\
  10.1103/physrevlett.100.073601}\BibitemShut {NoStop}%
\bibitem [{\citenamefont {Seshadreesan}\ \emph {et~al.}(2011)\citenamefont
  {Seshadreesan}, \citenamefont {Anisimov}, \citenamefont {Lee},\ and\
  \citenamefont {Dowling}}]{seshadreesan2011}%
  \BibitemOpen
  \bibfield  {author} {\bibinfo {author} {\bibfnamefont {Kaushik~P}\
  \bibnamefont {Seshadreesan}}, \bibinfo {author} {\bibfnamefont {Petr~M}\
  \bibnamefont {Anisimov}}, \bibinfo {author} {\bibfnamefont {Hwang}\
  \bibnamefont {Lee}}, \ and\ \bibinfo {author} {\bibfnamefont {Jonathan~P}\
  \bibnamefont {Dowling}},\ }\bibfield  {title} {\enquote {\bibinfo {title}
  {Parity detection achieves the heisenberg limit in interferometry with
  coherent mixed with squeezed vacuum light},}\ }\href {\doibase
  10.1088/1367-2630/13/8/083026} {\bibfield  {journal} {\bibinfo  {journal}
  {New Journal of Physics}\ }\textbf {\bibinfo {volume} {13}},\ \bibinfo
  {pages} {083026} (\bibinfo {year} {2011})}\BibitemShut {NoStop}%
\bibitem [{\citenamefont {Lawrie}\ \emph {et~al.}(2019)\citenamefont {Lawrie},
  \citenamefont {Lett}, \citenamefont {Marino},\ and\ \citenamefont
  {Pooser}}]{pooser2019}%
  \BibitemOpen
  \bibfield  {author} {\bibinfo {author} {\bibfnamefont {B.~J.}\ \bibnamefont
  {Lawrie}}, \bibinfo {author} {\bibfnamefont {P.~D.}\ \bibnamefont {Lett}},
  \bibinfo {author} {\bibfnamefont {A.~M.}\ \bibnamefont {Marino}}, \ and\
  \bibinfo {author} {\bibfnamefont {R.~C.}\ \bibnamefont {Pooser}},\ }\bibfield
   {title} {\enquote {\bibinfo {title} {Quantum sensing with squeezed light},}\
  }\href {\doibase 10.1021/acsphotonics.9b00250} {\bibfield  {journal}
  {\bibinfo  {journal} {{ACS} Photonics}\ }\textbf {\bibinfo {volume} {6}},\
  \bibinfo {pages} {1307--1318} (\bibinfo {year} {2019})}\BibitemShut {NoStop}%
\bibitem [{\citenamefont {Yurke}\ \emph {et~al.}(1986)\citenamefont {Yurke},
  \citenamefont {McCall},\ and\ \citenamefont {Klauder}}]{YMK1986}%
  \BibitemOpen
  \bibfield  {author} {\bibinfo {author} {\bibfnamefont {Bernard}\ \bibnamefont
  {Yurke}}, \bibinfo {author} {\bibfnamefont {Samuel~L.}\ \bibnamefont
  {McCall}}, \ and\ \bibinfo {author} {\bibfnamefont {John~R.}\ \bibnamefont
  {Klauder}},\ }\bibfield  {title} {\enquote {\bibinfo {title} {{SU}(2) and
  {SU}(1, 1) interferometers},}\ }\href {\doibase 10.1103/physreva.33.4033}
  {\bibfield  {journal} {\bibinfo  {journal} {Physical Review A}\ }\textbf
  {\bibinfo {volume} {33}},\ \bibinfo {pages} {4033--4054} (\bibinfo {year}
  {1986})}\BibitemShut {NoStop}%
\bibitem [{\citenamefont {Plick}\ \emph {et~al.}(2010)\citenamefont {Plick},
  \citenamefont {Dowling},\ and\ \citenamefont {Agarwal}}]{plick2010}%
  \BibitemOpen
  \bibfield  {author} {\bibinfo {author} {\bibfnamefont {William~N}\
  \bibnamefont {Plick}}, \bibinfo {author} {\bibfnamefont {Jonathan~P}\
  \bibnamefont {Dowling}}, \ and\ \bibinfo {author} {\bibfnamefont {Girish~S}\
  \bibnamefont {Agarwal}},\ }\bibfield  {title} {\enquote {\bibinfo {title}
  {Coherent-light-boosted, sub-shot noise, quantum interferometry},}\ }\href
  {\doibase 10.1088/1367-2630/12/8/083014} {\bibfield  {journal} {\bibinfo
  {journal} {New Journal of Physics}\ }\textbf {\bibinfo {volume} {12}},\
  \bibinfo {pages} {083014} (\bibinfo {year} {2010})}\BibitemShut {NoStop}%
\bibitem [{\citenamefont {Marino}\ \emph {et~al.}(2012)\citenamefont {Marino},
  \citenamefont {Trejo},\ and\ \citenamefont {Lett}}]{marino2012}%
  \BibitemOpen
  \bibfield  {author} {\bibinfo {author} {\bibfnamefont {A.~M.}\ \bibnamefont
  {Marino}}, \bibinfo {author} {\bibfnamefont {N.~V.~Corzo}\ \bibnamefont
  {Trejo}}, \ and\ \bibinfo {author} {\bibfnamefont {P.~D.}\ \bibnamefont
  {Lett}},\ }\bibfield  {title} {\enquote {\bibinfo {title} {Effect of losses
  on the performance of an {SU}(1, 1) interferometer},}\ }\href {\doibase
  10.1103/physreva.86.023844} {\bibfield  {journal} {\bibinfo  {journal}
  {Physical Review A}\ }\textbf {\bibinfo {volume} {86}} (\bibinfo {year}
  {2012}),\ 10.1103/physreva.86.023844}\BibitemShut {NoStop}%
\bibitem [{\citenamefont {Hudelist}\ \emph {et~al.}(2014)\citenamefont
  {Hudelist}, \citenamefont {Kong}, \citenamefont {Liu}, \citenamefont {Jing},
  \citenamefont {Ou},\ and\ \citenamefont {Zhang}}]{hudelist2014}%
  \BibitemOpen
  \bibfield  {author} {\bibinfo {author} {\bibfnamefont {F.}~\bibnamefont
  {Hudelist}}, \bibinfo {author} {\bibfnamefont {Jia}\ \bibnamefont {Kong}},
  \bibinfo {author} {\bibfnamefont {Cunjin}\ \bibnamefont {Liu}}, \bibinfo
  {author} {\bibfnamefont {Jietai}\ \bibnamefont {Jing}}, \bibinfo {author}
  {\bibfnamefont {Z.Y.}\ \bibnamefont {Ou}}, \ and\ \bibinfo {author}
  {\bibfnamefont {Weiping}\ \bibnamefont {Zhang}},\ }\bibfield  {title}
  {\enquote {\bibinfo {title} {Quantum metrology with parametric
  amplifier-based photon correlation interferometers},}\ }\href {\doibase
  10.1038/ncomms4049} {\bibfield  {journal} {\bibinfo  {journal} {Nature
  Communications}\ }\textbf {\bibinfo {volume} {5}} (\bibinfo {year} {2014}),\
  10.1038/ncomms4049}\BibitemShut {NoStop}%
\bibitem [{\citenamefont {Anisimov}\ \emph {et~al.}(2010)\citenamefont
  {Anisimov}, \citenamefont {Raterman}, \citenamefont {Chiruvelli},
  \citenamefont {Plick}, \citenamefont {Huver}, \citenamefont {Lee},\ and\
  \citenamefont {Dowling}}]{anisimov2010}%
  \BibitemOpen
  \bibfield  {author} {\bibinfo {author} {\bibfnamefont {Petr~M.}\ \bibnamefont
  {Anisimov}}, \bibinfo {author} {\bibfnamefont {Gretchen~M.}\ \bibnamefont
  {Raterman}}, \bibinfo {author} {\bibfnamefont {Aravind}\ \bibnamefont
  {Chiruvelli}}, \bibinfo {author} {\bibfnamefont {William~N.}\ \bibnamefont
  {Plick}}, \bibinfo {author} {\bibfnamefont {Sean~D.}\ \bibnamefont {Huver}},
  \bibinfo {author} {\bibfnamefont {Hwang}\ \bibnamefont {Lee}}, \ and\
  \bibinfo {author} {\bibfnamefont {Jonathan~P.}\ \bibnamefont {Dowling}},\
  }\bibfield  {title} {\enquote {\bibinfo {title} {Quantum metrology with
  two-mode squeezed vacuum: Parity detection beats the heisenberg limit},}\
  }\href {\doibase 10.1103/physrevlett.104.103602} {\bibfield  {journal}
  {\bibinfo  {journal} {Physical Review Letters}\ }\textbf {\bibinfo {volume}
  {104}} (\bibinfo {year} {2010}),\ 10.1103/physrevlett.104.103602}\BibitemShut
  {NoStop}%
\bibitem [{\citenamefont {Yuen}(1986)}]{Yuen1986}%
  \BibitemOpen
  \bibfield  {author} {\bibinfo {author} {\bibfnamefont {Horace~P.}\
  \bibnamefont {Yuen}},\ }\bibfield  {title} {\enquote {\bibinfo {title}
  {Generation, detection, and application of high-intensity
  photon-number-eigenstate fields},}\ }\href {\doibase
  10.1103/physrevlett.56.2176} {\bibfield  {journal} {\bibinfo  {journal}
  {Physical Review Letters}\ }\textbf {\bibinfo {volume} {56}},\ \bibinfo
  {pages} {2176--2179} (\bibinfo {year} {1986})}\BibitemShut {NoStop}%
\bibitem [{\citenamefont {Holland}\ and\ \citenamefont
  {Burnett}(1993)}]{holland1993}%
  \BibitemOpen
  \bibfield  {author} {\bibinfo {author} {\bibfnamefont {M.~J.}\ \bibnamefont
  {Holland}}\ and\ \bibinfo {author} {\bibfnamefont {K.}~\bibnamefont
  {Burnett}},\ }\bibfield  {title} {\enquote {\bibinfo {title} {Interferometric
  detection of optical phase shifts at the heisenberg limit},}\ }\href
  {\doibase 10.1103/physrevlett.71.1355} {\bibfield  {journal} {\bibinfo
  {journal} {Physical Review Letters}\ }\textbf {\bibinfo {volume} {71}},\
  \bibinfo {pages} {1355--1358} (\bibinfo {year} {1993})}\BibitemShut {NoStop}%
\bibitem [{\citenamefont {Han}\ and\ \citenamefont {Kim}(1998)}]{kim1998}%
  \BibitemOpen
  \bibfield  {author} {\bibinfo {author} {\bibfnamefont {D.}~\bibnamefont
  {Han}}\ and\ \bibinfo {author} {\bibfnamefont {Y.~S.}\ \bibnamefont {Kim}},\
  }\href {\doibase 10.48550/ARXIV.PHYSICS/9803017} {\enquote {\bibinfo {title}
  {Squeezed states as representations of symplectic groups},}\ } (\bibinfo
  {year} {1998})\BibitemShut {NoStop}%
\bibitem [{\citenamefont {Kuzmich}\ and\ \citenamefont
  {Mandel}(1998)}]{kuzmich1998}%
  \BibitemOpen
  \bibfield  {author} {\bibinfo {author} {\bibfnamefont {A}~\bibnamefont
  {Kuzmich}}\ and\ \bibinfo {author} {\bibfnamefont {L}~\bibnamefont
  {Mandel}},\ }\bibfield  {title} {\enquote {\bibinfo {title} {Sub-shot-noise
  interferometric measurements with two-photon states},}\ }\href {\doibase
  10.1088/1355-5111/10/3/008} {\bibfield  {journal} {\bibinfo  {journal}
  {Quantum and Semiclassical Optics: Journal of the European Optical Society
  Part B}\ }\textbf {\bibinfo {volume} {10}},\ \bibinfo {pages} {493--500}
  (\bibinfo {year} {1998})}\BibitemShut {NoStop}%
\bibitem [{\citenamefont {Berry}\ and\ \citenamefont
  {Wiseman}(2000)}]{berry2000}%
  \BibitemOpen
  \bibfield  {author} {\bibinfo {author} {\bibfnamefont {D.~W.}\ \bibnamefont
  {Berry}}\ and\ \bibinfo {author} {\bibfnamefont {H.~M.}\ \bibnamefont
  {Wiseman}},\ }\bibfield  {title} {\enquote {\bibinfo {title} {Optimal states
  and almost optimal adaptive measurements for quantum interferometry},}\
  }\href {\doibase 10.1103/physrevlett.85.5098} {\bibfield  {journal} {\bibinfo
   {journal} {Physical Review Letters}\ }\textbf {\bibinfo {volume} {85}},\
  \bibinfo {pages} {5098--5101} (\bibinfo {year} {2000})}\BibitemShut {NoStop}%
\bibitem [{\citenamefont {Campos}\ \emph {et~al.}(2003)\citenamefont {Campos},
  \citenamefont {Gerry},\ and\ \citenamefont {Benmoussa}}]{campos2003}%
  \BibitemOpen
  \bibfield  {author} {\bibinfo {author} {\bibfnamefont {R.~A.}\ \bibnamefont
  {Campos}}, \bibinfo {author} {\bibfnamefont {Christopher~C.}\ \bibnamefont
  {Gerry}}, \ and\ \bibinfo {author} {\bibfnamefont {A.}~\bibnamefont
  {Benmoussa}},\ }\bibfield  {title} {\enquote {\bibinfo {title} {Optical
  interferometry at the heisenberg limit with twin fock states and parity
  measurements},}\ }\href {\doibase 10.1103/physreva.68.023810} {\bibfield
  {journal} {\bibinfo  {journal} {Physical Review A}\ }\textbf {\bibinfo
  {volume} {68}} (\bibinfo {year} {2003}),\
  10.1103/physreva.68.023810}\BibitemShut {NoStop}%
\bibitem [{\citenamefont {Pezz\'e}\ and\ \citenamefont
  {Smerzi}(2006)}]{pezze2006}%
  \BibitemOpen
  \bibfield  {author} {\bibinfo {author} {\bibfnamefont {Luca}\ \bibnamefont
  {Pezz\'e}}\ and\ \bibinfo {author} {\bibfnamefont {Augusto}\ \bibnamefont
  {Smerzi}},\ }\bibfield  {title} {\enquote {\bibinfo {title} {Phase
  sensitivity of a mach-zehnder interferometer},}\ }\href {\doibase
  10.1103/PhysRevA.73.011801} {\bibfield  {journal} {\bibinfo  {journal} {Phys.
  Rev. A}\ }\textbf {\bibinfo {volume} {73}},\ \bibinfo {pages} {011801}
  (\bibinfo {year} {2006})}\BibitemShut {NoStop}%
\bibitem [{\citenamefont {Lee}\ \emph {et~al.}(2009)\citenamefont {Lee},
  \citenamefont {Huver}, \citenamefont {Lee}, \citenamefont {Kaplan},
  \citenamefont {McCracken}, \citenamefont {Min}, \citenamefont {Uskov},
  \citenamefont {Wildfeuer}, \citenamefont {Veronis},\ and\ \citenamefont
  {Dowling}}]{lee2009}%
  \BibitemOpen
  \bibfield  {author} {\bibinfo {author} {\bibfnamefont {Tae-Woo}\ \bibnamefont
  {Lee}}, \bibinfo {author} {\bibfnamefont {Sean~D.}\ \bibnamefont {Huver}},
  \bibinfo {author} {\bibfnamefont {Hwang}\ \bibnamefont {Lee}}, \bibinfo
  {author} {\bibfnamefont {Lev}\ \bibnamefont {Kaplan}}, \bibinfo {author}
  {\bibfnamefont {Steven~B.}\ \bibnamefont {McCracken}}, \bibinfo {author}
  {\bibfnamefont {Changjun}\ \bibnamefont {Min}}, \bibinfo {author}
  {\bibfnamefont {Dmitry~B.}\ \bibnamefont {Uskov}}, \bibinfo {author}
  {\bibfnamefont {Christoph~F.}\ \bibnamefont {Wildfeuer}}, \bibinfo {author}
  {\bibfnamefont {Georgios}\ \bibnamefont {Veronis}}, \ and\ \bibinfo {author}
  {\bibfnamefont {Jonathan~P.}\ \bibnamefont {Dowling}},\ }\bibfield  {title}
  {\enquote {\bibinfo {title} {Optimization of quantum interferometric
  metrological sensors in the presence of photon loss},}\ }\href {\doibase
  10.1103/physreva.80.063803} {\bibfield  {journal} {\bibinfo  {journal}
  {Physical Review A}\ }\textbf {\bibinfo {volume} {80}} (\bibinfo {year}
  {2009}),\ 10.1103/physreva.80.063803}\BibitemShut {NoStop}%
\bibitem [{\citenamefont {Higgins}\ \emph {et~al.}(2007)\citenamefont
  {Higgins}, \citenamefont {Berry}, \citenamefont {Bartlett}, \citenamefont
  {Wiseman},\ and\ \citenamefont {Pryde}}]{higgins2007}%
  \BibitemOpen
  \bibfield  {author} {\bibinfo {author} {\bibfnamefont {B.~L.}\ \bibnamefont
  {Higgins}}, \bibinfo {author} {\bibfnamefont {D.~W.}\ \bibnamefont {Berry}},
  \bibinfo {author} {\bibfnamefont {S.~D.}\ \bibnamefont {Bartlett}}, \bibinfo
  {author} {\bibfnamefont {H.~M.}\ \bibnamefont {Wiseman}}, \ and\ \bibinfo
  {author} {\bibfnamefont {G.~J.}\ \bibnamefont {Pryde}},\ }\bibfield  {title}
  {\enquote {\bibinfo {title} {Entanglement-free heisenberg-limited phase
  estimation},}\ }\href {\doibase 10.1038/nature06257} {\bibfield  {journal}
  {\bibinfo  {journal} {Nature}\ }\textbf {\bibinfo {volume} {450}},\ \bibinfo
  {pages} {393--396} (\bibinfo {year} {2007})}\BibitemShut {NoStop}%
\bibitem [{\citenamefont {Nagata}\ \emph {et~al.}(2007)\citenamefont {Nagata},
  \citenamefont {Okamoto}, \citenamefont {O{\textquotesingle}Brien},
  \citenamefont {Sasaki},\ and\ \citenamefont {Takeuchi}}]{nagata2007}%
  \BibitemOpen
  \bibfield  {author} {\bibinfo {author} {\bibfnamefont {Tomohisa}\
  \bibnamefont {Nagata}}, \bibinfo {author} {\bibfnamefont {Ryo}\ \bibnamefont
  {Okamoto}}, \bibinfo {author} {\bibfnamefont {Jeremy~L.}\ \bibnamefont
  {O{\textquotesingle}Brien}}, \bibinfo {author} {\bibfnamefont {Keiji}\
  \bibnamefont {Sasaki}}, \ and\ \bibinfo {author} {\bibfnamefont {Shigeki}\
  \bibnamefont {Takeuchi}},\ }\bibfield  {title} {\enquote {\bibinfo {title}
  {Beating the standard quantum limit with four-entangled photons},}\ }\href
  {\doibase 10.1126/science.1138007} {\bibfield  {journal} {\bibinfo  {journal}
  {Science}\ }\textbf {\bibinfo {volume} {316}},\ \bibinfo {pages} {726--729}
  (\bibinfo {year} {2007})}\BibitemShut {NoStop}%
\bibitem [{\citenamefont {Hillery}\ and\ \citenamefont
  {Mlodinow}(1993)}]{hillery1993}%
  \BibitemOpen
  \bibfield  {author} {\bibinfo {author} {\bibfnamefont {Mark}\ \bibnamefont
  {Hillery}}\ and\ \bibinfo {author} {\bibfnamefont {Leonard}\ \bibnamefont
  {Mlodinow}},\ }\bibfield  {title} {\enquote {\bibinfo {title}
  {Interferometers and minimum-uncertainty states},}\ }\href {\doibase
  10.1103/PhysRevA.48.1548} {\bibfield  {journal} {\bibinfo  {journal} {Phys.
  Rev. A}\ }\textbf {\bibinfo {volume} {48}},\ \bibinfo {pages} {1548--1558}
  (\bibinfo {year} {1993})}\BibitemShut {NoStop}%
\bibitem [{\citenamefont {Brif}\ and\ \citenamefont {Mann}(1996)}]{brif1996}%
  \BibitemOpen
  \bibfield  {author} {\bibinfo {author} {\bibfnamefont {C.}~\bibnamefont
  {Brif}}\ and\ \bibinfo {author} {\bibfnamefont {A.}~\bibnamefont {Mann}},\
  }\bibfield  {title} {\enquote {\bibinfo {title} {Nonclassical interferometry
  with intelligent light},}\ }\href {\doibase 10.1103/physreva.54.4505}
  {\bibfield  {journal} {\bibinfo  {journal} {Physical Review A}\ }\textbf
  {\bibinfo {volume} {54}},\ \bibinfo {pages} {4505--4518} (\bibinfo {year}
  {1996})}\BibitemShut {NoStop}%
\bibitem [{\citenamefont {Lee}\ \emph {et~al.}(2002)\citenamefont {Lee},
  \citenamefont {Kok},\ and\ \citenamefont {Dowling}}]{lee2002}%
  \BibitemOpen
  \bibfield  {author} {\bibinfo {author} {\bibfnamefont {Hwang}\ \bibnamefont
  {Lee}}, \bibinfo {author} {\bibfnamefont {Pieter}\ \bibnamefont {Kok}}, \
  and\ \bibinfo {author} {\bibfnamefont {Jonathan~P.}\ \bibnamefont
  {Dowling}},\ }\bibfield  {title} {\enquote {\bibinfo {title} {A quantum
  rosetta stone for interferometry},}\ }\href {\doibase
  10.1080/0950034021000011536} {\bibfield  {journal} {\bibinfo  {journal}
  {Journal of Modern Optics}\ }\textbf {\bibinfo {volume} {49}},\ \bibinfo
  {pages} {2325--2338} (\bibinfo {year} {2002})}\BibitemShut {NoStop}%
\bibitem [{\citenamefont {Resch}\ \emph {et~al.}(2007)\citenamefont {Resch},
  \citenamefont {Pregnell}, \citenamefont {Prevedel}, \citenamefont
  {Gilchrist}, \citenamefont {Pryde}, \citenamefont {O'Brien},\ and\
  \citenamefont {White}}]{resch2007}%
  \BibitemOpen
  \bibfield  {author} {\bibinfo {author} {\bibfnamefont {K.~J.}\ \bibnamefont
  {Resch}}, \bibinfo {author} {\bibfnamefont {K.~L.}\ \bibnamefont {Pregnell}},
  \bibinfo {author} {\bibfnamefont {R.}~\bibnamefont {Prevedel}}, \bibinfo
  {author} {\bibfnamefont {A.}~\bibnamefont {Gilchrist}}, \bibinfo {author}
  {\bibfnamefont {G.~J.}\ \bibnamefont {Pryde}}, \bibinfo {author}
  {\bibfnamefont {J.~L.}\ \bibnamefont {O'Brien}}, \ and\ \bibinfo {author}
  {\bibfnamefont {A.~G.}\ \bibnamefont {White}},\ }\bibfield  {title} {\enquote
  {\bibinfo {title} {Time-reversal and super-resolving phase measurements},}\
  }\href {\doibase 10.1103/PhysRevLett.98.223601} {\bibfield  {journal}
  {\bibinfo  {journal} {Phys. Rev. Lett.}\ }\textbf {\bibinfo {volume} {98}},\
  \bibinfo {pages} {223601} (\bibinfo {year} {2007})}\BibitemShut {NoStop}%
\bibitem [{\citenamefont {Afek}\ \emph {et~al.}(2010)\citenamefont {Afek},
  \citenamefont {Ambar},\ and\ \citenamefont {Silberberg}}]{afek2010}%
  \BibitemOpen
  \bibfield  {author} {\bibinfo {author} {\bibfnamefont {Itai}\ \bibnamefont
  {Afek}}, \bibinfo {author} {\bibfnamefont {Oron}\ \bibnamefont {Ambar}}, \
  and\ \bibinfo {author} {\bibfnamefont {Yaron}\ \bibnamefont {Silberberg}},\
  }\bibfield  {title} {\enquote {\bibinfo {title} {High-{NOON} states by mixing
  quantum and classical light},}\ }\href {\doibase 10.1126/science.1188172}
  {\bibfield  {journal} {\bibinfo  {journal} {Science}\ }\textbf {\bibinfo
  {volume} {328}},\ \bibinfo {pages} {879--881} (\bibinfo {year}
  {2010})}\BibitemShut {NoStop}%
\bibitem [{\citenamefont {Birrittella}\ \emph {et~al.}(2012)\citenamefont
  {Birrittella}, \citenamefont {Mimih},\ and\ \citenamefont
  {Gerry}}]{birrittella2012}%
  \BibitemOpen
  \bibfield  {author} {\bibinfo {author} {\bibfnamefont {Richard}\ \bibnamefont
  {Birrittella}}, \bibinfo {author} {\bibfnamefont {Jihane}\ \bibnamefont
  {Mimih}}, \ and\ \bibinfo {author} {\bibfnamefont {Christopher~C.}\
  \bibnamefont {Gerry}},\ }\bibfield  {title} {\enquote {\bibinfo {title}
  {Multiphoton quantum interference at a beam splitter and the approach to
  heisenberg-limited interferometry},}\ }\href {\doibase
  10.1103/physreva.86.063828} {\bibfield  {journal} {\bibinfo  {journal}
  {Physical Review A}\ }\textbf {\bibinfo {volume} {86}} (\bibinfo {year}
  {2012}),\ 10.1103/physreva.86.063828}\BibitemShut {NoStop}%
\bibitem [{\citenamefont {Carranza}\ and\ \citenamefont
  {Gerry}(2012)}]{carranza2012}%
  \BibitemOpen
  \bibfield  {author} {\bibinfo {author} {\bibfnamefont {Raul}\ \bibnamefont
  {Carranza}}\ and\ \bibinfo {author} {\bibfnamefont {Christopher~C.}\
  \bibnamefont {Gerry}},\ }\bibfield  {title} {\enquote {\bibinfo {title}
  {Photon-subtracted two-mode squeezed vacuum states and applications to
  quantum optical interferometry},}\ }\href {\doibase 10.1364/josab.29.002581}
  {\bibfield  {journal} {\bibinfo  {journal} {Journal of the Optical Society of
  America B}\ }\textbf {\bibinfo {volume} {29}},\ \bibinfo {pages} {2581}
  (\bibinfo {year} {2012})}\BibitemShut {NoStop}%
\bibitem [{\citenamefont {Zhang}\ \emph {et~al.}(2021)\citenamefont {Zhang},
  \citenamefont {Ye}, \citenamefont {Wei}, \citenamefont {Xia}, \citenamefont
  {Chang}, \citenamefont {Liao},\ and\ \citenamefont {Hu}}]{zhang2021}%
  \BibitemOpen
  \bibfield  {author} {\bibinfo {author} {\bibfnamefont {Huan}\ \bibnamefont
  {Zhang}}, \bibinfo {author} {\bibfnamefont {Wei}\ \bibnamefont {Ye}},
  \bibinfo {author} {\bibfnamefont {Chaoping}\ \bibnamefont {Wei}}, \bibinfo
  {author} {\bibfnamefont {Ying}\ \bibnamefont {Xia}}, \bibinfo {author}
  {\bibfnamefont {Shoukang}\ \bibnamefont {Chang}}, \bibinfo {author}
  {\bibfnamefont {Zeyang}\ \bibnamefont {Liao}}, \ and\ \bibinfo {author}
  {\bibfnamefont {Liyun}\ \bibnamefont {Hu}},\ }\bibfield  {title} {\enquote
  {\bibinfo {title} {Improved phase sensitivity in a quantum optical
  interferometer based on multiphoton catalytic two-mode squeezed vacuum
  states},}\ }\href {\doibase 10.1103/physreva.103.013705} {\bibfield
  {journal} {\bibinfo  {journal} {Physical Review A}\ }\textbf {\bibinfo
  {volume} {103}} (\bibinfo {year} {2021}),\
  10.1103/physreva.103.013705}\BibitemShut {NoStop}%
\bibitem [{\citenamefont {Kumar}\ \emph {et~al.}(2022)\citenamefont {Kumar},
  \citenamefont {Rishabh},\ and\ \citenamefont {Arora}}]{kumar2022}%
  \BibitemOpen
  \bibfield  {author} {\bibinfo {author} {\bibfnamefont {Chandan}\ \bibnamefont
  {Kumar}}, \bibinfo {author} {\bibnamefont {Rishabh}}, \ and\ \bibinfo
  {author} {\bibfnamefont {Shikhar}\ \bibnamefont {Arora}},\ }\bibfield
  {title} {\enquote {\bibinfo {title} {Realistic non-gaussian-operation scheme
  in parity-detection-based mach-zehnder quantum interferometry},}\ }\href
  {\doibase 10.1103/physreva.105.052437} {\bibfield  {journal} {\bibinfo
  {journal} {Physical Review A}\ }\textbf {\bibinfo {volume} {105}} (\bibinfo
  {year} {2022}),\ 10.1103/physreva.105.052437}\BibitemShut {NoStop}%
\bibitem [{\citenamefont {Xiao}\ \emph {et~al.}(2020)\citenamefont {Xiao},
  \citenamefont {Matekole}, \citenamefont {Zhao}, \citenamefont {Zeng},
  \citenamefont {Dowling},\ and\ \citenamefont {Lee}}]{xiao2020}%
  \BibitemOpen
  \bibfield  {author} {\bibinfo {author} {\bibfnamefont {Xiao-Qi}\ \bibnamefont
  {Xiao}}, \bibinfo {author} {\bibfnamefont {Elisha~S.}\ \bibnamefont
  {Matekole}}, \bibinfo {author} {\bibfnamefont {Jiankang}\ \bibnamefont
  {Zhao}}, \bibinfo {author} {\bibfnamefont {Guihua}\ \bibnamefont {Zeng}},
  \bibinfo {author} {\bibfnamefont {Jonathan~P.}\ \bibnamefont {Dowling}}, \
  and\ \bibinfo {author} {\bibfnamefont {Hwang}\ \bibnamefont {Lee}},\
  }\bibfield  {title} {\enquote {\bibinfo {title} {Enhanced phase estimation
  with coherently boosted two-mode squeezed beams and its application to
  optical gyroscopes},}\ }\href {\doibase 10.1103/PhysRevA.102.022614}
  {\bibfield  {journal} {\bibinfo  {journal} {Phys. Rev. A}\ }\textbf {\bibinfo
  {volume} {102}},\ \bibinfo {pages} {022614} (\bibinfo {year}
  {2020})}\BibitemShut {NoStop}%
\bibitem [{\citenamefont {Weedbrook}\ \emph {et~al.}(2012)\citenamefont
  {Weedbrook}, \citenamefont {Pirandola}, \citenamefont
  {Garc{\'\i}a-Patr{\'o}n}, \citenamefont {Cerf}, \citenamefont {Ralph},
  \citenamefont {Shapiro},\ and\ \citenamefont
  {Lloyd}}]{weedbrook2012gaussian}%
  \BibitemOpen
  \bibfield  {author} {\bibinfo {author} {\bibfnamefont {Christian}\
  \bibnamefont {Weedbrook}}, \bibinfo {author} {\bibfnamefont {Stefano}\
  \bibnamefont {Pirandola}}, \bibinfo {author} {\bibfnamefont {Ra{\'u}l}\
  \bibnamefont {Garc{\'\i}a-Patr{\'o}n}}, \bibinfo {author} {\bibfnamefont
  {Nicolas~J}\ \bibnamefont {Cerf}}, \bibinfo {author} {\bibfnamefont
  {Timothy~C}\ \bibnamefont {Ralph}}, \bibinfo {author} {\bibfnamefont
  {Jeffrey~H}\ \bibnamefont {Shapiro}}, \ and\ \bibinfo {author} {\bibfnamefont
  {Seth}\ \bibnamefont {Lloyd}},\ }\bibfield  {title} {\enquote {\bibinfo
  {title} {Gaussian quantum information},}\ }\href
  {https://journals.aps.org/rmp/abstract/10.1103/RevModPhys.84.621} {\bibfield
  {journal} {\bibinfo  {journal} {Reviews of Modern Physics}\ }\textbf
  {\bibinfo {volume} {84}},\ \bibinfo {pages} {621} (\bibinfo {year}
  {2012})}\BibitemShut {NoStop}%
\bibitem [{\citenamefont {Gard}(2016)}]{gard2016advances}%
  \BibitemOpen
  \bibfield  {author} {\bibinfo {author} {\bibfnamefont {Bryan~Tomas}\
  \bibnamefont {Gard}},\ }\href
  {https://digitalcommons.lsu.edu/cgi/viewcontent.cgi?article=3088&context=gradschool_dissertations}
  {\emph {\bibinfo {title} {Advances in quantum metrology: Continuous variables
  in phase space}}}\ (\bibinfo  {publisher} {Louisiana State University and
  Agricultural \& Mechanical College},\ \bibinfo {year} {2016})\BibitemShut
  {NoStop}%
\bibitem [{\citenamefont {Gard}\ \emph {et~al.}(2017)\citenamefont {Gard},
  \citenamefont {You}, \citenamefont {Mishra}, \citenamefont {Singh},
  \citenamefont {Lee}, \citenamefont {Corbitt},\ and\ \citenamefont
  {Dowling}}]{gard2017nearly}%
  \BibitemOpen
  \bibfield  {author} {\bibinfo {author} {\bibfnamefont {Bryan~T}\ \bibnamefont
  {Gard}}, \bibinfo {author} {\bibfnamefont {Chenglong}\ \bibnamefont {You}},
  \bibinfo {author} {\bibfnamefont {Devendra~K}\ \bibnamefont {Mishra}},
  \bibinfo {author} {\bibfnamefont {Robinjeet}\ \bibnamefont {Singh}}, \bibinfo
  {author} {\bibfnamefont {Hwang}\ \bibnamefont {Lee}}, \bibinfo {author}
  {\bibfnamefont {Thomas~R}\ \bibnamefont {Corbitt}}, \ and\ \bibinfo {author}
  {\bibfnamefont {Jonathan~P}\ \bibnamefont {Dowling}},\ }\bibfield  {title}
  {\enquote {\bibinfo {title} {Nearly optimal measurement schemes in a noisy
  mach-zehnder interferometer with coherent and squeezed vacuum},}\ }\href
  {https://epjquantumtechnology.springeropen.com/articles/10.1140/epjqt/s40507-017-0058-8}
  {\bibfield  {journal} {\bibinfo  {journal} {EPJ Quantum Technology}\ }\textbf
  {\bibinfo {volume} {4}},\ \bibinfo {pages} {1--13} (\bibinfo {year}
  {2017})}\BibitemShut {NoStop}%
\end{thebibliography}%

\end{document}